\documentclass[journal=ancac3,manuscript=article]{achemso}
\usepackage{achemso}
\usepackage{graphicx}
\usepackage{capt-of}
\usepackage[english]{babel}
\usepackage[utf8]{inputenc}
\usepackage{siunitx}
\usepackage{sidecap}
\usepackage{subfigure}
\usepackage{capt-of}
\usepackage{amssymb}
\usepackage{xcolor,xspace,colortbl,rotating}
\usepackage{amsfonts}
\usepackage[version=3]{mhchem}
\title{{Magnetic domain engineering in antiferromagnetic CuMnAs and \ce{Mn2Au} devices}}
\author{Sonka Reimers}
\affiliation{School of Physics and Astronomy, University of Nottingham, Nottingham NG7 2RD, United Kingdom}
      \email{sreimers@uni-mainz.de}
     \alsoaffiliation{Institut f\"ur Physik, Johannes Gutenberg-Universit\"at Mainz, 55099 Mainz, Germany}
     \alsoaffiliation{Diamond Light Source, Chilton OX11 0DE, United Kingdom}
    \author{Olena Gomonay}
	\affiliation{Institut f\"ur Physik, Johannes Gutenberg-Universit\"at Mainz, 55099 Mainz, Germany}
    \author{Oliver J. Amin}
    \affiliation{School of Physics and Astronomy, University of Nottingham, Nottingham NG7 2RD, United Kingdom}
    \author{Filip Krizek}
    \affiliation{Institute of Physics, Czech Academy of Sciences, 162 00 Praha 6, Czech Republic}
    \author{Luke X. Barton}
    \affiliation{School of Physics and Astronomy, University of Nottingham, Nottingham NG7 2RD, United Kingdom}
    \author{Yaryna Lytvynenko}
	\affiliation{Institut f\"ur Physik, Johannes Gutenberg-Universit\"at Mainz, 55099 Mainz, Germany}
 \alsoaffiliation{Institute of Magnetism of the NAS of Ukraine and MES of Ukraine, 03142 Kyiv, Ukraine}
    \author{Stuart F. Poole}
    \affiliation{School of Physics and Astronomy, University of Nottingham, Nottingham NG7 2RD, United Kingdom}       
    \author{Vit Nov\'{a}k}
    \affiliation{Institute of Physics, Czech Academy of Sciences, 162 00 Praha 6, Czech Republic}
    \author{Richard P. Campion}
    \affiliation{School of Physics and Astronomy, University of Nottingham, Nottingham NG7 2RD, United Kingdom}
   	\author{Francesco Maccherozzi}
    \affiliation{Diamond Light Source, Chilton OX11 0DE, United Kingdom}
    \author{Dina Carbone}
    \affiliation{MaxIV Laboratory, 224 84 Lund, Sweden}
    \author{Alexander Bj{\"o}rling}
    \affiliation{MaxIV Laboratory, 224 84 Lund, Sweden}
    \author{Yuran Niu}
    \affiliation{MaxIV Laboratory, 224 84 Lund, Sweden}
        \author{Evangelos Golias}
    \affiliation{MaxIV Laboratory, 224 84 Lund, Sweden}
     \author{Dominik Kriegner}
    \affiliation{Institute of Solid State and Materials Physics, Technical University Dresden, 01069 Dresden, Germany}
    \alsoaffiliation{Institute of Physics, Czech Academy of Sciences, 162 00 Praha 6, Czech Republic}
	\author{Jairo Sinova}
	\affiliation{Institut f\"ur Physik, Johannes Gutenberg-Universit\"at Mainz, 55099 Mainz, Germany}
    \author{Mathias Kläui}
	\affiliation{Institut f\"ur Physik, Johannes Gutenberg-Universit\"at Mainz, 55099 Mainz, Germany}
 \alsoaffiliation{Center for Quantum Spintronics, Norwegian University of Science and Technology NTNU, Trondheim, Norway}
      \author{Martin Jourdan}
	\affiliation{Institut f\"ur Physik, Johannes Gutenberg-Universit\"at Mainz, 55099 Mainz, Germany}
    \author{Sarnjeet S. Dhesi}
    \affiliation{Diamond Light Source, Chilton OX11 0DE, United Kingdom}
    \author{Kevin W. Edmonds}
    \affiliation{School of Physics and Astronomy, University of Nottingham, Nottingham NG7 2RD, United Kingdom}
    \author{Peter Wadley}
    \affiliation{School of Physics and Astronomy, University of Nottingham, Nottingham NG7 2RD, United Kingdom}

\begin{document}
\maketitle

% Author: Please give full first and last names for authors and include * after the name of all corresponding authors
% Dedication
%\dedication{Optional dedication here. If no dedication is required, please leave blank}
% Affiliations: Please provide adacemic titles (Prof. or Dr.) for all authors where applicable, and include an institutional email address for all corresponding authors

% Keywords: Please provide a minimum of three and a maximum of seven keywords, separated by commas
% Abstract should be written in the present tense and impersonal style (i.e., avoid we), and be at most 200 words long
\newpage
\begin{abstract}

Antiferromagnetic materials hold potential for use in spintronic devices with fast operation frequencies and field robustness. Despite the rapid progress in proof-of-principle functionality in recent years, there has been a notable lack of understanding of antiferromagnetic domain formation and manipulation, which translates to either incomplete or non-scalable control of the magnetic order.  
Here, we demonstrate simple and functional ways of influencing the domain structure in CuMnAs and \ce{Mn2Au}, two key materials of antiferromagnetic spintronics research, using device patterning and strain engineering. Comparing x-ray microscopy data from two different materials, we reveal the key parameters dictating domain formation in antiferromagnetic devices and show how the non-trivial interaction of magnetostriction, substrate clamping and edge anisotropy leads to specific equilibrium domain configurations. More specifically, we observe that patterned edges have a significant impact on the magnetic anisotropy and domain structure over long distances, and we propose a theoretical model that relates short-range edge anisotropy and long-range magnetoelastic interactions. The principles invoked are of general applicability to the domain formation and engineering in antiferromagnetic thin films at large, which will pave the way towards realizing truly functional antiferromagnetic devices.

% Antiferromagnetic materials hold potential for use in spintronic devices with fast operation frequencies and field robustness and recent years have seen a rapid progress in proof-of-principle functionality. Yet, there has been a lack in the understanding of antiferromagnetic domain formation and manipulation, which translates to either incomplete control of the magnetic
% order or the utilization of non-scalable control. Here, we demonstrate simple and functional ways of influencing the domain structure in CuMnAs and Mn2Au, two key materials in antiferromagnetic spintronics research, using device patterning and strain engineering. 
% Using data from two different materials, we reveal the key parameters which dictate domain formation in devices and show, how the non-trivial interaction of magnetostriction, substrate clamping, and edge anisotropy leads to specific equilibrium domain configurations in devices. These principles that are applicable generally to domain formation and engineering in antiferromagnetic thin films.

\end{abstract}

\newpage
\section{Introduction}
Antiferromagnetic (AF) spintronics has the potential to be a breakthrough technology for data storage in terms of speed, scaling, and robustness. 
Following the theoretical prediction of efficient manipulation of AF order with electrical currents \cite{Zelezny2014}, research has focused on proof-of-principle functionality. Thus, rapid progress has been made in this respect both experimentally and theoretically \cite{Chen2023, Qin2023, Wadley2018, Kosub2017, Wadley16, Review22}.
Yet, most cases either show incomplete control of the AF order, lack long-term stability or require large current pulses and significant heating or even non-scalable approaches, such as the application of large magnetic fields. To overcome these limitations and realise truly functional devices from antiferromagnets requires an in-depth understanding of antiferromagnetic domain formation and a precise tuning of the local magnetic anisotropy in devices, which to date remains mostly elusive for fully compensated antiferromagnets.

%In this context, current-pulse-induced N\'eel vector switching is widely studied for both insulating and metallic antiferromagnets \cite{Review22}. 
%The magnetic anisotropy is a key parameter for electrical switching experiments and for future applications, as it defines the barrier to be overcome and stabilizes the N\'eel vector orientation.
In ferromagnetic devices, domain formation is largely governed by the minimization of magnetic stray fields, which makes the magnetic anisotropy sensitive to the shape of the device\cite{Kittel_Domains}.
However, magnetic stray fields are entirely absent in the bulk of fully compensated antiferromagnets, so that a-priori long-range shape-induced phenomena may not be expected.

Nonetheless, it has been reported for devices fabricated from antiferromagnetic oxide films that the domain structure becomes sensitive to the device shape, when the dimensions are on the micrometer scale.
Studies with \ce{LaFeO3} have revealed both short-range and long-range effects \cite{Shapes_LFO, Folven2010, Folven2012} with short-range effects related to an edge anisotropy, which leads to a local alignment of the AF spin axes parallel or perpendicular to the lithographic edge. This is mediated only by direct AF exchange over distances up to the typical AF domain size. Effects over longer ranges were attributed to magnetoelastic interactions. 
Also for NiO/Pt heterostructures, it has recently been shown that minimization of elastic energy has a primary role in the domain configuration \cite{Hendrik2022}. In NiO, magnetoelasticity originates from exchange-coupling and hence is known to be large.

In the antiferromagnetic metals CuMnAs and \ce{Mn2Au}, studied here, magnetoelastic coupling can only arise from the relativistic spin-orbit coupling, which is typically several orders of magnitude smaller than exchange coupling. Nevertheless, as we show in this work, patterned edges significantly influence the domain structure and magnetic anisotropy also in CuMnAs and \ce{Mn2Au} thin films.
Tetragonal CuMnAs and \ce{Mn2Au} are amongst the most promising material candidates for application in AF spintronics, because they are the only materials discovered for which electrical currents can rotate the Néel vector via so-called N\'eel spin-orbit torques (NSOTs)\cite{Wadley16, Wadley2018, Baldrati2019, Lytvynenko2022, Godinho2018}. Insight into the mechanism of domain formation and anisotropy in these materials is thus directly relevant for AF spintronics research and potential future application. 

To understand the effect of patterning on the domain structure, we study the AF domains in simple geometries fabricated from CuMnAs and \ce{Mn2Au} films using x-ray magnetic linear dichroism photoemission electron microscopy (XMlD-PEEM).

Our experimental data reveal a spatial variation of the local magnetic anisotropy over several micrometers, which we measure as a change of the domain wall widths in CuMnAs and as a change of the average domain population in the \ce{Mn2Au} films. This microscopic investigation of the anisotropy in AF structures gives unique insight into the mechanisms which underpin AF domain formation and anisotropy changes in patterned structures. We show that the spatial variation of the anisotropy and the characteristic domain structures are directly related to the spatial distribution of strain in the samples, which we model using magnetoelastic charges.  
We find that short-range edge induced anisotropy arising at the patterned edges can induce long-range effects due to magnetoelastic interactions competing with the film-substrate clamping. 
By comparing CuMnAs and \ce{Mn2Au}, we disentangle the contributions to the local anisotropy and show how in patterned structures the same mechanisms result in different stable AF domain morphologies depending on the material parameters. Our description can be generally applied to domain formation in devices fabricated from epitaxial AF thin films grown on a nonmagnetic substrates, the most relevant geometry for spintronic application. This study gives a crucial contribution towards closing the gap between proof-of-principle experiments and truly functional AF spintronic devices.

\subsection*{Samples}
Here we study epitaxial (001)-oriented CuMnAs and \ce{Mn2Au} films grown on nonmagnetic substrates. 
The films have a tetragonal crystal structure and exhibit antiferromagnetic ordering at room temperature, with magnetic moments located on Mn sites. The magnetic moments align ferromagnetically in the (001) plane, with antiferromagnetic coupling between neighbouring planes \cite{Barthem13, wadley13}. Both materials exhibit large out-of-plane magnetocrystalline anisotropies, confining the AF spin axis in-plane. The in-plane magnetic anisotropies, which are much smaller, favour two mutually orthogonal magnetic easy axes along the $\langle 110 \rangle$ crystallographic directions that match the in-plane symmetry of the substrates.
 Thus, from a magnetoelastic point of view, the samples have the same geometry, but the size of the magnetocrystalline anisotropies vary considerably. For \ce{Mn2Au}, states with spin alignment along the $\langle 110\rangle$ and $\langle 100 \rangle$ crystallographic directions are separated by an energy barrier larger than $ \SI{1.8}{\micro eV~ per}$ formula unit (f.u.)\cite{Bommana21}.
 For  CuMnAs, experiment and theory indicate near-degeneracy of the spin axis direction in the $(ab)$ plane, with  the energy difference between the $\langle 110\rangle$ and $\langle 100 \rangle$ crystallographic directions $< \SI{1}{\micro eV~ per~}$f.u.~close to the resolution limit of the calculation\cite{Hills2015} and a preference for the $\langle 110\rangle$ directions (magnetic easy axes hereafter) in \SI{50}{nm} films as studied here. The out-of-plane anisotropy barrier is far larger in both materials, with \SI{2}{meV~ per~}f.u.~in \ce{Mn2Au} \cite{Barthem13}, and \SI{127}{\micro eV~ per} f.u.~in CuMnAs\cite{Hills2015}.  

\subsection*{AF domain imaging}
 The AF domain structures are imaged by photoemission electron microscopy (PEEM) with sensitivity to the axis of the N\'eel vector (AF spin axis hereafter) due to the x-ray magnetic linear dichroism (XMLD) effect. For x-ray polarization along an in-plane high-symmetry  crystallographic axis, maximum contrast is achieved between domains aligned parallel and perpendicular to the x-ray polarization direction. The XMLD-PEEM imaging of AF domains has been established for both materials in previous experiments \cite{wadley17,Bodnar2018, Bommana21}.

\section{Results and Discussion}

First, we discuss the patterning-induced AF domain configurations using CuMnAs as an example and introduce the magnetoelastic model. Later in this manuscript we show how this model can be applied to the second investigated antiferromagnet \ce{Mn2Au}, which has a much larger magnetocrystalline anisotropy.

The model considers three contributions to the total energy: magnetic energy that contains exchange energy  $W_\mathrm{ex}$ and magnetocrystalline anisotropy $W_\mathrm{mc}$, 
surface energy at the patterned edge $W_\mathrm{edge}$, referred to hereafter as edge energy, and destressing energy at the film-substrate interface $W_\mathrm{destr}$, which is the elastic energy due to magnetostriction of the antiferromagnet and film-substrate clamping. We distinguish between epitaxial (or growth-induced) strain and spontaneous incompatibility strain. The epitaxial strain is isotropic and hence does not break the symmetry between the two magnetic easy axes, and can be neglected in many cases. Only directly at the edge, this strain can be relaxed anisotropically over approximately the distance of the film thickness. This is one possible origin of edge anisotropy. With the term incompatibility, we refer to the additional spontaneous strain that emerges due to magnetoelastic coupling when comparing the non-magnetic state with the magnetic state. This is anisotropic and therefore can affect AF domain formation. Further information can be found in the Methods section.

\subsection*{Effect of patterning in CuMnAs}
 \begin{figure}[hbt]
 \centering
 \includegraphics[width = \textwidth]{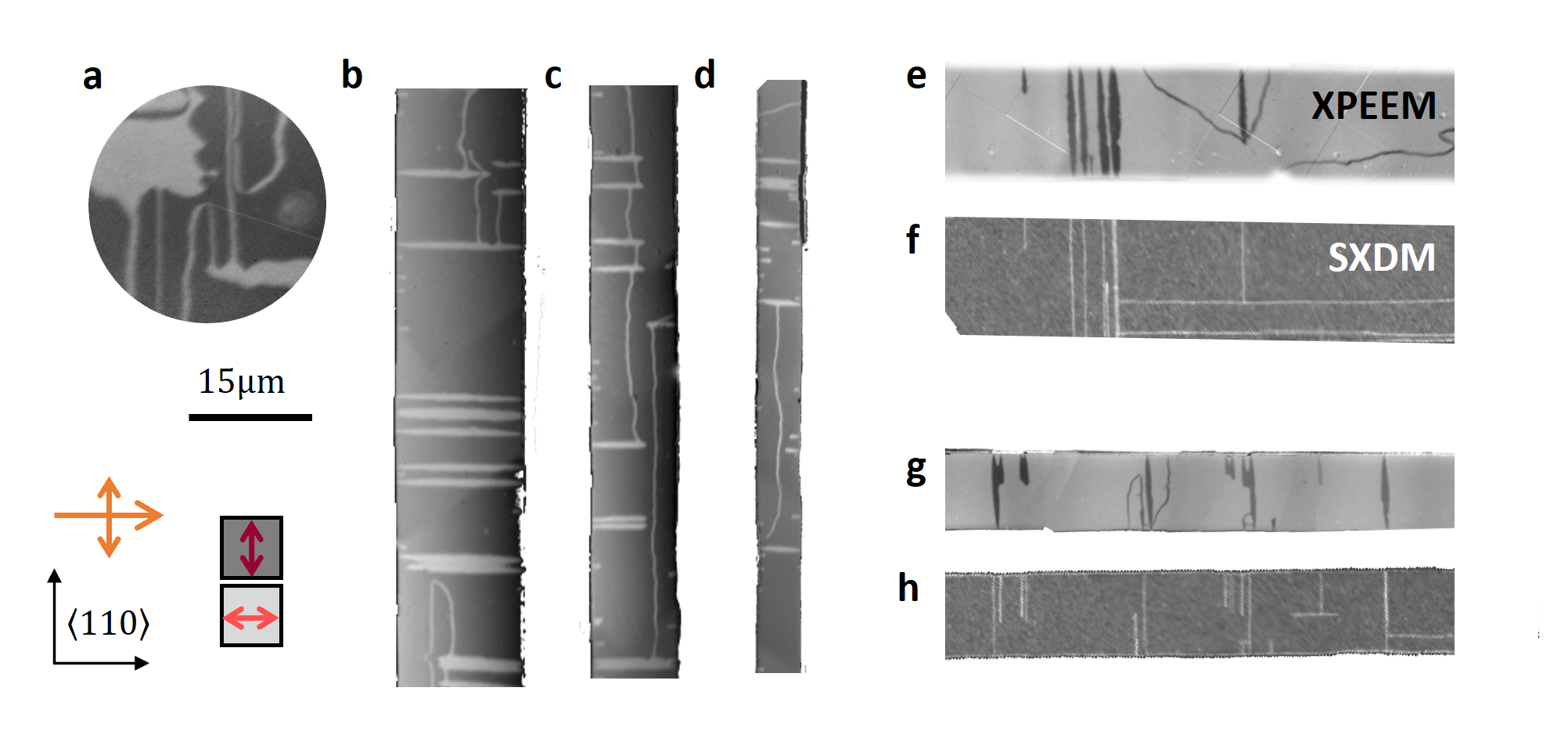}
  \caption{Effect of patterning along a magnetic easy axis in CuMnAs.  \textbf{a)} AF domains in a non-patterned area imaged by XMLD-PEEM. The orange arrows show the direction and polarization of the incident x-ray beam, which yields sensitivity to the spin axis as shown by the red double-headed arrows. \textbf{b}-\textbf{d)}  AF domain images in \SI{15}{\micro \meter}, \SI{10}{\micro \meter} and \SI{5}{\micro \meter} wide bars along an easy axes. \textbf{e)} AF domain image in a \SI{15}{\micro \meter} wide bar with orthogonal orientation. \textbf{f)} scanning x-ray diffraction microscopy (SXDM) image of the same bar showing the microtwin configuration. \textbf{g},\textbf{h)} same as \textbf{e}-\textbf{f)} but for a \SI{10}{\micro \meter} wide bar. }
    \label{fig:CuMnAsEA}
\end{figure}
High quality epitaxial CuMnAs films, of \SI{50}{nm} thickness, are grown by molecular beam epitaxy or GaP(001) substrates with a mismatch between film and substrate of about $\sim \SI{0.3}{\percent}$. The films are grown fully strained.
Epitaxial strain is relaxed locally in the vicinity of specific crystallographic nanoscale defects referred to as microtwins \cite{Krizek2020}.
Non-patterned (as-grown) samples of the films studied here consist of two types of AF domains with mutually orthogonal spin axes, along the $\langle 110 \rangle$ crystallographic directions of the films. Previous work has shown that the domain morphology and size are governed by nanoscale microtwin defects \cite{Reimers2022}. 
 The microtwins terminate at the surface as characteristic lines along the crystallographic $\langle 110 \rangle$ directions and locally pin the AF spin axis parallel to their direction. The strain field surrounding such a twin defect can stabilize \SI{180}{\degree} domain walls due to magnetostriction\cite{Reimers2022}. 
For the sake of simplicity, we will refer to the $\langle 110\rangle$ directions as the magnetic easy axis, and the $\langle 1o0\rangle$ directions as the magnetic hard axis, for these films.

Structures with different geometric shapes are fabricated using photolithography and chemical wet etching from a layer with low microtwin density and typical domain sizes beyond \SI{10}{\micro \meter ^2}.  A representative example of the domain structure in a non-patterned area is shown in \textbf{Figure \ref{fig:CuMnAsEA}\,a}. 

\subsubsection*{Patterning along a magnetic easy axis}
We first consider the simplest device geometry consisting of bars oriented along a magnetic easy axis.
Figure~\ref{fig:CuMnAsEA} compares the AF domain structures in bars of different widths (\textbf{b}-\textbf{g}) to the domain structure in a non-patterned area (\textbf{a}).
A strong effect of the patterning on the AF domain population and morphology is observed. In a non-patterned area, the population of orthogonal domain types is almost equal, whereas in the bars,
the AF spin axis is aligned mostly parallel to the long edge: the horizontal bars (Fig.~\ref{fig:CuMnAsEA}\,\textbf{e,g}) show a prevailing presence of horizontal spin axis (light areas), whereas the vertical bars (Fig.~\ref{fig:CuMnAsEA}\,\textbf{b-d})) appear mostly dark.
Domains with orthogonal spin axis are observed only as characteristic lens-shaped domains aligned perpendicular to the edges. Direct comparison to scanning x-ray diffraction microscopy (SXDM) images of the microtwin structure (panels \textbf{f,h}) show that each lens-shaped domain corresponds to a microtwin defect which is located at the centre of the lens-shaped domain. Microtwin-free areas are fully aligned parallel to the edge of the bar. 
We have observed the stabilization of the AF spin axis parallel to the edge across the entire widths of the bar, even in the largest \SI{25}{\micro \meter} wide structures.

In addition to the lens-shaped domains, \SI{180}{\degree} domain walls are observed. These \SI{180}{\degree} domain walls are characteristic for uniaxial anisotropy. They appear as narrow, undulating lines in the XMLD-PEEM images and often run parallel to the long side of the bar. However, when terminating at an edge, they locally align perpendicular to the edge, similar to what has been observed for \SI{180}{\degree} domain walls at the surfaces of uniaxial \ce{Cr2O3} single crystals \cite{Hedrich21}.   

\subsubsection*{Patterning along a magnetic hard axis}
\begin{SCfigure}[][hbt]
 \centering
  \includegraphics[width = 0.5\textwidth]{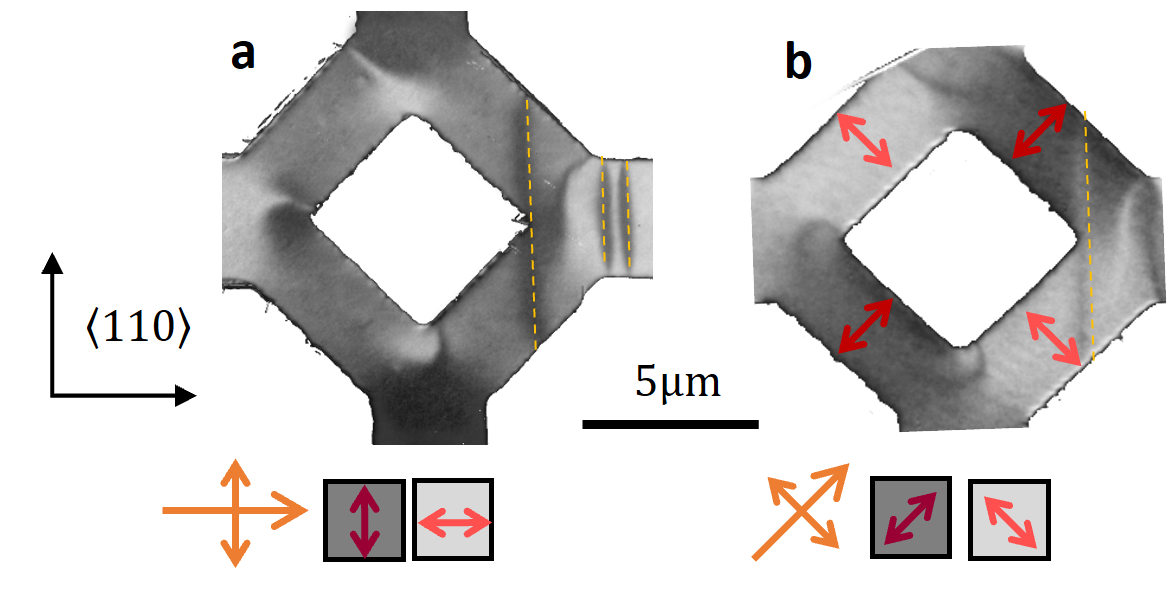}
  \caption{Effect of patterning along a magnetic hard axis in CuMnAs.  \textbf{a)} AF domains in a square-shaped, almost microtwin-free device imaged with x-ray polarization along a CuMnAs  $\langle 110 \rangle$ direction, as indicated by the double-headed orange arrows. The yellow dashed lines mark the positions of microtwin defects. \textbf{b)}  Same device, but imaged with x-ray polarization imaged with x-ray polarization along a CuMnAs  $\langle 100 \rangle$ direction, which reveals that the AF spin axis aligns perpendicular to the edge.} \label{fig:CuMnAsHA}
\end{SCfigure}

In the case of patterned edges along a magnetic hard axis (\textbf{Figure~\ref{fig:CuMnAsHA}}), the XMLD-PEEM images reveal that the AF spin axis aligns \textit{perpendicular} to the edge. This is opposite to what was observed for edges aligned with the magnetic easy axes. Thus, the patterning-induced anisotropy exceeds the intrinsic magnetocrystalline anisotropy of CuMnAs.  
In the vicinity of microtwin-defects, \textit{e.g.} at the right corner of the device in panels \textbf{a,b}, the competition of the effect of microtwins and the patterned edge leads to a frustrated domain structure. 

\subsubsection*{Description within the magnetoelastic model}
To describe domain formation in CuMnAs theoretically, we consider the case of a large edge anisotropy and large exchange energy compared to the destressing energy. 
In this scenario, the AF spin axis is fixed at the edges. As exchange energy dominates over destressing energy, the entire bar can end up in a single domain state dictated by the edge anisotropy. This is parallel to the bar for edges along a magnetic easy axis and perpendicular for edges along a magnetic hard axis.
 This kind of dependence of the sign of the edge anisotropy on the patterning direction has also been observed in LFO (see Ref. \cite{Shapes_LFO}). It clearly shows that in antiferromagnets edge anisotropy does not arise from the minimization of small magnetic stray fields, but can have different origins. It can arise, for example, from the different chemical environment of the surface atoms (\textit{e.g.}~oxidation) or from strain relaxation.

\begin{figure}[hbt]
   \centering
  \includegraphics[width = 0.69\textwidth]{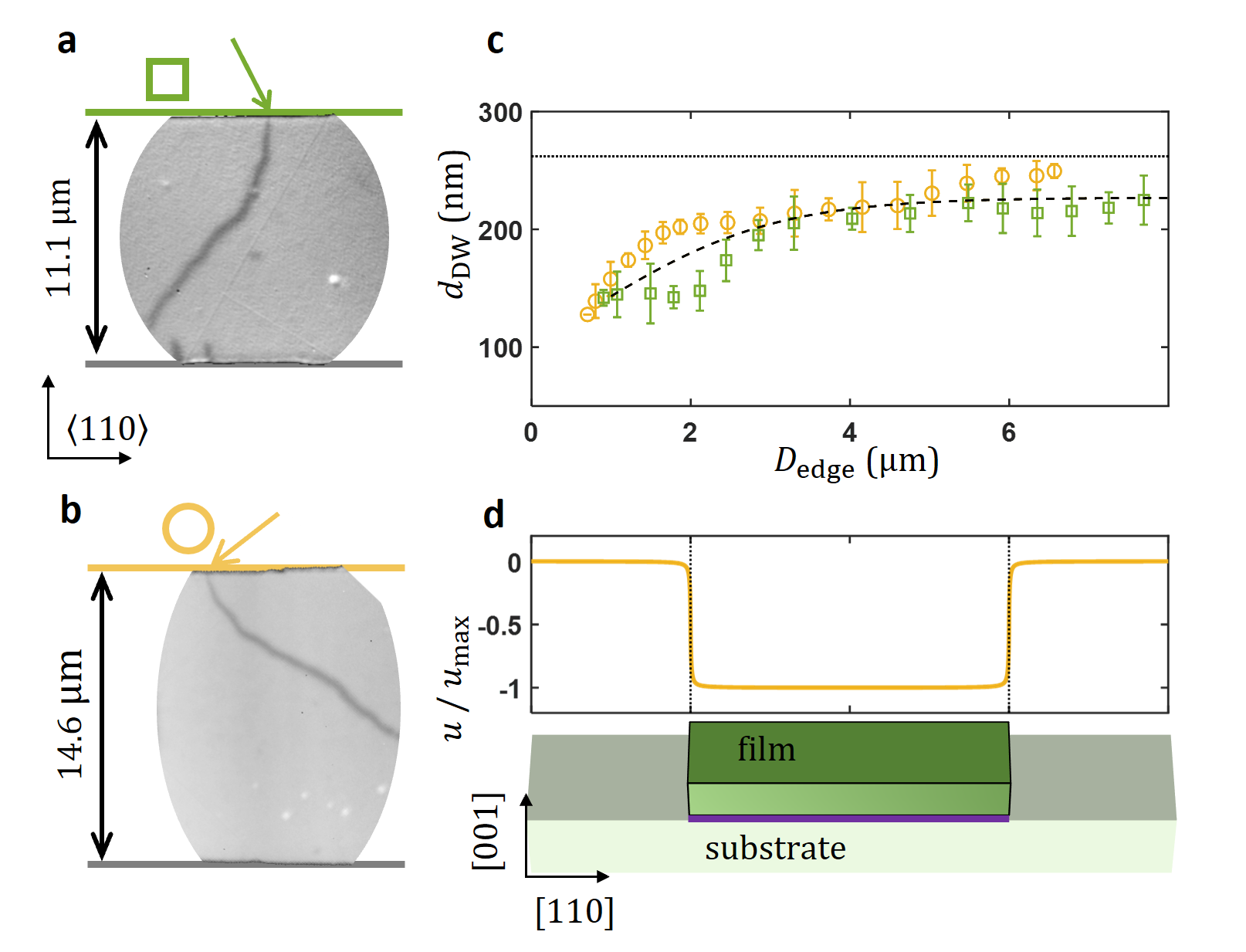}
    \caption{Gradient anisotropy in \ce{CuMnAs} bars. \textbf{a)},\textbf{b)} \SI{180}{\degree} DWs in a CuMnAs bars. \textbf{c)} Dependence of the widths of the DWs in panels \textbf{a},\textbf{b} on the distance to the edge, measured with respect to the colored edge. Green and yellow datapoints correspond to the images in panels \textbf{a} and \textbf{b}. Dashed line: simulated dependence of the DW width. Dotted line: DW width measured in non-patterned areas of the sample. 
   \textbf{d)} Schematic illustrating the origin of the effect. Purple sheet: magnetoelastic charge density causing the destressing field $u$ shown in the inset (yellow curve). This field opposes the single domain state and reduces the anisotropy towards the centre of the bar.}
    \label{fig:MechanismCuMnAs}
\end{figure}

For further insight into the mechanism, we study the spatial variation of the anisotropy, in particular its dependence on the distance to the patterned edges. 
The widths of \SI{180}{\degree} domain walls (DWs) act as a local probe of the magnetic anisotropy, since they scale as $1/\sqrt{K}$ with the anisotropy constant $K$\cite{hubert1998}. 
 Two representative examples of domain walls which terminate at edges are shown in \textbf{Figure~\ref{fig:MechanismCuMnAs}\,a} and \textbf{b}. Upon close inspection, it can be seen that the domain walls are considerably narrower in the direct vicinity of the edge. This can be clearly seen in a plot of the domain wall widths as a function of distance to the edge (panel \textbf{c}). The data show that the change occurs continuously over several micrometers. The domain wall widths reduce to almost half their value ($\sim $\SI{120}{\nano \meter}) in the vicinity of the edges compared to the centre of the bars (\SI{200}{\nano \meter} - \SI{250}{\nano \meter}). Even in the centre, they remain below the widths measured in non-patterned areas of approximately \SI{260}{nm} in this sample (shown by the dotted line).

This shows that  the anisotropy is not just altered directly at the edge, but over long-ranges. To understand this long-range effect, the elastic effect of the alignment of the AF spin axis needs to be considered: the alignment of the spin axis with the edge leads to a deformation of the film. This, vice versa, creates a preference for AF domains aligned accordingly, \textit{i.e.}~an additional uniaxial anisotropy and can explain, for example, the presence of \SI{180}{\degree} domain walls even in biaxial magnetic systems\cite{hubert1998}. 
Yet, the corresponding lattice deformation of the film is incompatible with the non-deformed substrate.

Mathematically, we model this incompatibility with an elastic charge density at the film-substrate interface, illustrated by the purple sheet in Figure~\ref{fig:MechanismCuMnAs}\,\textbf{d}.   
A single domain configuration corresponds to a uniform charge density. This charge density creates a long-range ``Coulomb-like" elastic field in film and substrate, similar to the electric field of a uniformly charged finite plate. It is largest in the centre of the bar and counteracts the patterning induced effect. Hence the anisotropy is largest at the edge and reduces towards the centre of the bar. This translates into a characteristic dependence of the domain wall widths as a function of distance to the edge (see Methods section), shown by the dashed line in Figure~\ref{fig:MechanismCuMnAs}\,c,  which is an averaged fit of the datasets. 
The good agreement of the model with the experimental data supports the validity of our theoretical description. Deviations of the data from the theoretical curve can be ascribed to local crystallographic defects. 

We have shown that the AF domain formation is governed by the competition of edge anisotropy, exchange and destressing energy. In the CuMnAs structures, the destressing energy is small and the final state is largely governed by edge anisotropy and exchange energy. The effect of the destressing energy manifests itself as the broadening of the domain walls towards the centre of the bars.
The competition between the edge anisotropy and the destressing field becomes more obvious in the \ce{Mn2Au} films studied and discussed next.
 
\subsection*{Patterning effects in \ce{Mn2Au}}
The (001) oriented \ce{Mn2Au} films are \SI{40}{nm} thick and grown epitaxially on Ta(001)/Mo(001) double buffer layers on MgO(001) substrates \cite{Lytvynenko2022}. As mentioned above, \ce{Mn2Au}, like CuMnAs, has two mutually orthogonal, equivalent magnetic easy axes but a much larger magnetocrystalline anisotropy.
Both magnetocrystalline anisotropy and magnetoelasticity originate from spin-orbit interaction  and therefore magnetoelasticity can also be expected to be considerably larger in \ce{Mn2Au} than in CuMnAs, so that the destressing energy becomes more relevant.

 Without patterning, the epitaxial \ce{Mn2Au}(001) thin films show a multidomain AF domain state, with two types of domains aligned with the two magnetic easy axes and equipartitial distribution of the two domain types. A representative example is shown in \textbf{Figure~\ref{fig:Mn2Au}\,a}. The domain size is in the order of \SI{1}{\micro \meter}, considerably smaller compared to the CuMnAs samples discussed above. Consistent with our assumption, a smaller domain size is expected for systems in which the destressing energy is more relevant. No evidence of morphological structures or crystallographic defects on the same lengthscale as the magnetic domains was found.
The width of the domain walls in \ce{Mn2Au} is less than \SI{80}{nm}, close to the resolution limit of the XMLD-PEEM technique, much smaller than in CuMnAs, where the domain wall widths can vary significantly depending on the local defect structure  \cite{Reimers2022}. This shows the larger magnetocrystalline anisotropy of \ce{Mn2Au}. 

To study the effect of patterning on this type of domain structure, devices with similar bar geometries are fabricated using photolithography and \ce{Ar+} ion beam milling. 
Figure~\ref{fig:Mn2Au} compares the AF domain structure in patterned bars with different orientations to the non-patterned case.

\subsubsection*{Patterning along a magnetic easy axis}
\begin{figure}[hbt] 
     \centering
     \includegraphics[width = \textwidth]{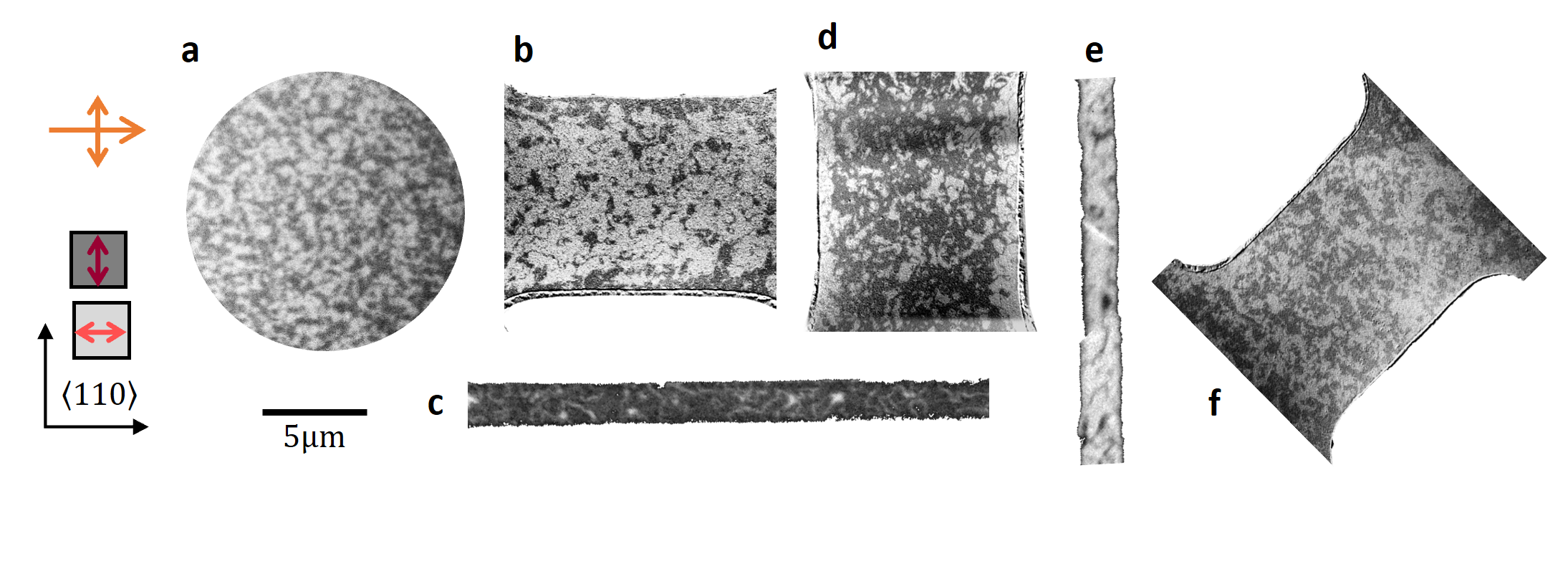}
    \caption{Effect of patterning in \ce{Mn2Au}. \textbf{a)} in a non-patterned area. \textbf{b)},\textbf{c)} in a \SI{10}{\micro \meter} and a \SI{2}{\micro \meter}  wide bar along a magnetic easy axes. \textbf{d)},\textbf{e)} Same as \textbf{b},\textbf{c} but for bars with orthogonal orientation. \textbf{f)} AF domains in a \SI{10}{\micro \meter} wide bar along a magnetic hard axis. The orange arrows show the direction and polarization of the incident x-ray beam.}
    \label{fig:Mn2Au}
\end{figure}

As in the case of CuMnAs, the bars oriented along magnetic easy axes (Figure~\ref{fig:Mn2Au}\,\textbf{b-e}) show a pronounced effect of the microlithography on the AF domain structure. However, in contrast to CuMnAs, the competing effects of the edge anisotropy and the destressing field can be observed directly: in the \SI{10}{\micro \meter} wide bars (panel \textbf{b}) the AF spin axis is aligned perpendicular to the edge in a narrow $ \sim \SI{0.5}{\micro \meter} -  \SI{1}{\micro \meter}$ near-edge region. This orientation is stabilized across the entire bar, if the width of the bar approaches the typical domain size, as in the narrow \SI{2}{\micro \meter} wide bars. In these, only one type of domain and \SI{180}{\degree} domain walls are observed.
In the wider bars away from the near-edge region domains with their AF spin axis along both perpendicular easy axes are observed, but domains with AF spin axis parallel to the edges clearly dominate the domain population. 
This indicates that the effects of the destressing field are dominating.

\subsubsection*{Patterning along a magnetic hard axis}
Figure~\ref{fig:Mn2Au}\,\textbf{f} shows the AF domains in a bar patterned along a magnetic hard axis. In this device, no effect of the patterning on the domain structure can be observed. The only detectable variation compared to non-patterned areas occurs in the corners of the device, \textit{i.e.}~where the edge is locally aligned with a magnetic easy axis.  
Hence, in \ce{Mn2Au}, the edge anisotropy is not sufficiently strong to overcome the intrinsic anisotropy and cannot force the AF spin axis onto a hard axis of the magnetocrystalline anisotropy.

\subsubsection*{Description within the magnetoelastic model}
\begin{figure}[hbt]
    \centering
   \includegraphics[width = \textwidth]{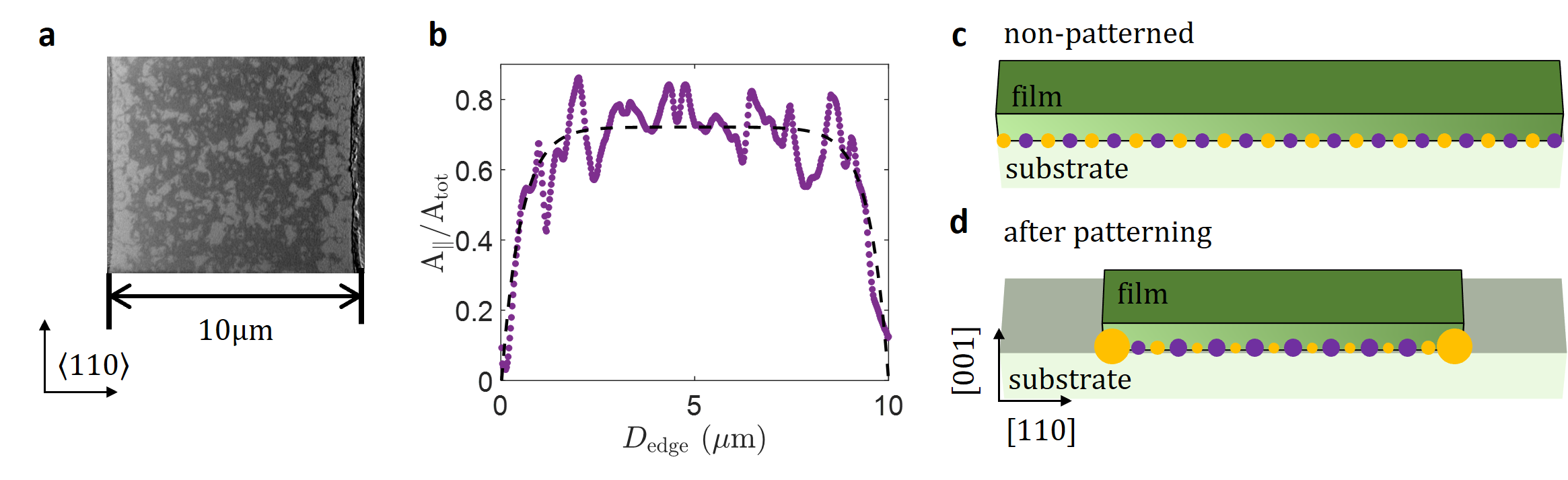}
    \caption{Gradient anisotropy in \ce{Mn2Au} bars. \textbf{a)} AF domains in a \SI{10}{\micro \meter} wide \ce{Mn2Au} bar. \textbf{b)} Average domain population as a function of distance to the left edge. The purple dots are experimental data, the black dashed line is a theoretical simulation. \textbf{c}) Magnetoelastic charge configuration before patterning. Purple and yellow correspond to opposite charge. \textbf{d}) Magnetoelastic charge configuration after patterning. edge anisotropy imposes alignment of the AF spin axis at the edges, creating an imbalance of charges. This increases the destressing energy, so that the film reacts by the expansion of domains with opposite charge in the centre. }
    \label{fig:MechanismMn2Au}
\end{figure}

\ce{Mn2Au} is an example where the magnetocrystalline anisotropy and the destressing energy are relatively large. Consequently, as-grown, unpatterned samples show a multi-domain state with equipartitial population of the two domain types, which minimizes destressing energy and patterning-induced anisotropy cannot fully rotate the AF spin axis, but only affects the relative population of domains. 
Thus, patterned edges along a magnetic hard axis have no measurable effect of the AF domain structure. 
In narrow (\SI{2}{\micro m}) bars along the magnetic easy axis, the AF spin axis is almost fully aligned perpendicular to the edge, which can be explained by the edge anisotropy and direct exchange alone; the \SI{180}{\degree} domain walls are of kinetic origin. However as the destressing energy is larger than in CuMnAs, this alignment is not maintained over the complete width of the wider (\SI{10}{\micro \meter}) bars.
In these wider bars, the AF domain population in the centre of the bars is reversed compared to the near-edge region. 
The variation of the domain population as a function of distance to the edge is shown in \textbf{Figure~\ref{fig:MechanismMn2Au}}. The nonequipartitial distribution of domains is observed across the entire width of the bars, which exceed the typical AF domain size by an order of magnitude. This is clear evidence for an elastic origin of the effect and can be described within our model as follows.

The alignment of the AF spin axis of the near-edge region due to edge anisotropy deforms the lattice of the film, so that  the average strain is nonzero at the edges and creates additional strain due to incompatibility (modelled as magnetoelastic charges of one sign). This increases the destressing energy compared to a state with zero average strain.  
Hence, \ce{Mn2Au} reacts by creating domains with orthogonal spin axis  (charges of opposite sign). Mathematically, we can model the incompatibility with a magnetoelastic charge density at the interface. The sign and size of the charge density is proportional to the ratio of the two domain types, hence the centre and the boundary layer have opposite sign. We then start with an initial guess of the distribution of the domains, calculate the charge density and corresponding energy and iteratively approach a state which minimizes the energy, corresponding to a stable AF domain configuration. Consistent with experiment, the model predicts a sharp rise of the domain population over a short distance and almost a plateau in the central area in good agreement with the experimental data.

\subsection*{Relevance for current induced switching experiments}
CuMnAs and \ce{Mn2Au} are two of the most important conducting materials for AF spintronics. Most experimental and theoretical studies with these materials are concerned with the manipulation of their AF order via NSOTs.
Yet, the behaviour of the AF domain structure is directly related to the AF anisotropy, the  equilibrium domain structure as well as several potential metastable states. As we show here, these  depend on the shape of the device. Thus, our findings are of great relevance for the interpretation of previous work on AF domain manipulation and can be used to optimize devices for future experiments.

Our data show that there is an equilibrium domain configuration in patterned structures, \textit{i.e.}~in potential AF spintronics devices, which is different from the as-grown state.
In the context of switching between two orthogonal orientations of the N\'eel vector, as originally proposed for AF magnetic random access memory (MRAM), this raises the question, if the switched states can be long-term stable and if the AF domain structure relaxes to the equilibrium configuration by thermal activation. Here we note two things: firstly, in CuMnAs, neither electrical current pulses nor magnetic field have been able to induce long-term stable switching by \SI{90}{\degree} of a large fraction of the active area of a device at room temperature. On the contrary, the switching is characterized by significant decay \cite{KamilTHz, Khalid_Switching, Wang2020}. This can be understood by considering the existence of a ground state that is dictated by patterning in combination with the low magnetocrystalline anisotropy. Secondly, for \ce{Mn2Au}, long-term stable switching of a large area of a device at room-temperature has been demonstrated and no relaxation of electrical switching at room temperature is observed \cite{Lytvynenko2022}. This is because the domain walls in this material are narrower and more strongly pinned. The lack of relaxation at room-temperature is consistent with the fact that the AF domain structure is altered during the patterning process, because the \ce{Ar+} ion beam milling used significantly heats the sample. This helps to overcome the domain wall pinning barriers required to achieve the new equilibrium configuration, whereas at room-temperature, the pinning barriers are sufficiently large to prevent relaxation of the current-induced meta-stable state 
This example illustrates that applications rely on a fine tuning of the anisotropy and that knowledge of the effect of patterning is crucial in order to design suitable device geometries. 
For example, for applications which rely on orthogonal switching of the N\'eel vector between two magnetic easy axis, edges along the magnetic easy axes lead to significant pinning, thus limit the device efficiency. Consistently, in \mbox{reference~\cite{Lytvynenko2022}}, long-term stable and reversible switching of effectively the entire active area of a \ce{Mn2Au} device was shown with an ``easy-edge-free'' device geometry.
Recent work also suggested applications beyond AF MRAM, including novel computing based on complex AF textures such as vortices and AF merons \cite{Radaelli2021}. Of the two materials studied there, these are most likely to be realizable in CuMnAs devices with low magnetocrystalline anisotropy, in which the competition of the edge effect and microtwin defects can be used to engineer a frustrated anisotropy landscape. This can be achieved by patterning edges along the magnetic hard axis. Indeed, such a geometry was used in \mbox{reference~\cite{OllieVortex}}, where AF merons were electrically generated.
In contrast, an enhanced uniaxial anisotropy might be beneficial for devices based on the motion of \SI{180}{\degree} domain walls, as could be envisioned for an AF race-track memory.

\section{Conclusion}
Microlithography has been shown to have a significant impact on the AF domains and magnetic anisotropy in both CuMnAs and \ce{Mn2Au} films. When edges are patterned, an additional anisotropy is induced, which results in a preferred alignment of the AF spin axis. This patterning-induced anisotropy can surpass the bulk magnetocrystalline anisotropy in materials like CuMnAs, where the intrinsic anisotropy is small. However, even in materials with larger intrinsic anisotropy, such as \ce{Mn2Au}, the patterning-induced anisotropy has a significant impact on the AF domain structure. In devices, this leads to pronounced gradients of the anisotropy that can be directly measured as a change of \SI{180}{\degree} domain wall widths in CuMnAs, or inferred from the average domain population in \ce{Mn2Au}. Patterned edges, which are inevitable in devices, can therefore be utilized as a tool to tune the anisotropy and domain structure for both applications and experiments involving these widely investigated materials.
By comparing data from both materials, a model of domain formation in AF devices has been developed, which allows for the disentanglement of the different contributions that govern the AF domain structure and the local magnetic anisotropy. For instance, the anisotropy gradients are shown to be governed by the spatial distribution of strain. The model establishes a relationship between short-range edge anisotropy and long-range magnetostrictive effects via magnetoelastic charges, in analogy to electrostatics. The principles of edge anisotropy, magnetostriction and anisotropic strain relaxation at patterned edges are generic to all antiferromagnetic films grown on non-magnetic substrates. Therefore, the model is generally applicable to antiferromagnetic devices.
The understanding of the mechanisms that govern domain formation and the local anisotropy in such structures forms the basis to precisely tune them towards specific applications. This will facilitate the realization of functional and efficient spintronic devices.

\section{Experimental Section}
\subsection*{Material growth}
The \SI{50}{\nano \meter} CuMnAs(001) films were grown by molecular beam epitaxy on a GaP buffer layer on GaP(001) substrates at \SI{210}{\celsius}. The films were capped with a \SI{3}{\nano \meter} Al layer to prevent surface oxidation. Details of the MBE-growth can be found in \mbox{reference~\cite{Krizek2020}}.
Ex-situ x-ray diffraction (XRD) diffraction measurements confirmed the tetragonal crystal structure of the layer, with the epitaxial relationship CuMnAs(001)$[100]||\mathrm{GaP(001)}[110]$. XRD measurements and scanning transmission electron microscopy (STEM) suggest that the films are grown fully strained with epitaxial strain relaxed only in the vicinity of specific crystallographic nanoscale defects, referred to as microtwins \cite{Krizek2020}.

The \SI{40}{nm} \ce{Mn2Au}(001) films were prepared by radio-frequency magnetron sputtering (rf-sputtering). The \ce{Mn2Au}(001) films were deposited on MgO(001) substrates with \SI{20}{nm} Mo(001) and \SI{16}{nm} Ta(001) double buffer layers and capped with \SI{2}{nm} \ce{Si3N4}. 
The  \ce{Mn2Au} layers were deposited at approximately $\SI{500}{\celsius}$ and subsequently annealed for \SI{75}{minutes} at $\SI{700}{\celsius}$. 
For further information see \mbox{reference~\cite{jourdan2015epitaxial}}.
The tetragonal structure and crystalline quality of the films was confirmed with ex-situ x-ray diffraction, establishing the \ce{Mn2Au}(001)$[100]||\mathrm{MgO(001)}[110]$ epitaxial relationship. 

In the main text, the crystallographic axes always refer to the CuMnAs\slash \ce{Mn2Au} films and not to the substrates.

\subsection*{Device fabrication}
The CuMnAs devices were fabricated by optical lithography and chemical wet etching. 

The \ce{Mn2Au} devices were fabricated by optical lithography and \ce{Ar+} ion beam milling with a continuous \ce{Ar+} current. The etching process removes the entire \ce{Mn2Au} and buffer layers and part of the substrate. 

\subsection*{XMLD-PEEM imaging}
The XMLD-PEEM measurements were performed on beamline I06 at \texttt{Diamond Light Source} and at the MAXPEEM beamline at MAX IV. On I06 at Diamond, the x-ray beam was incident at a grazing angle, forming an angle of \SI{16.5}{\degree} with the sample surface. Magnetic contrast was obtained from the difference in the absorption signal measured by XPEEM at the peak and the minimum of the Mn $\mathrm{L}_{2,3}$ XMLD spectrum.  
At the MAXPEEM beamline, the x-ray beam has normal incidence on the sample surface. Magnetic contrast was obtained from the difference of the absorption measured by XPEEM for two mutually perpendicular in-plane polarizations for a fixed photon energy at the maximum of the Mn $\mathrm{L}_{2,3}$ XMLD spectrum.
The different measurement configurations only affect the absolute scale of the signal, thus are irrelevant after image normalization. 

All measurements were performed at room-temperature and with approximately \SI{50}{nm} spatial resolution. 

The dependencies of the XMLD on the crystal orientation of CuMnAs was determined previously using an exchange-coupled Fe/CuMnAs  bilayer \cite{wadley17}. It was shown that the XMLD spectrum has a similar shape but opposite sign for $\vec{E}||[100]$ and $\vec{E}||[110]$.
For \ce{Mn2Au}, the dependence of the XMLD for $\vec{E}||[110]$ was established from \ce{Mn2Au} samples which have been exposed to an in-plane \SI{30}{T} magnetic field. This induces a spin-flop transition and leads to approximately \SI{80}{\percent} remanent AF spin axis orientation perpendicular to the field \cite{Sapozhnik18, Sapozhnik17}.

On beamline I06, images were acquired with a LEEM III microscope operated at \SI{12}{keV} for the CuMnAs samples and at \SI{15}{keV} for the \ce{Mn2Au} samples. On the MAXPEEM beamline at MAX IV, the LEEM III microscope was operated at \SI{20}{keV}.

\subsection*{Scanning x-ray diffraction microscopy (SXDM)}
SXDM imaging of microtwin patterns in CuMnAs devices was performed at the \texttt{NanoMAX} beamline at \texttt{MAX IV Laboratory}. The x-ray beam with an energy of \SI{10}{keV} was focused to a lateral diameter of \SI{100}{\nano \meter} onto the sample onto the sample, aligned to satisfy a Bragg condition, and rastered across the sample surface in a 2D mesh at different angles around the Bragg peak. The 3D reciprocal space maps thus obtained provided information about structural features of the sample (see \mbox{reference~\cite{Reimers2022}})and rastered across the sample surface in an $xy$-mesh during a scan.
 As discussed in \mbox{reference~\cite{Reimers2022}}, the microtwins create specific features
(wings) at Bragg reflection peaks.
This allows to map the microtwin configuration by plotting the intensity measured in s against the sample position during a scan.  These intensity maps were obtained at the CuMnAs (003) peak of the sample at an angle $\Delta \Theta = \pm \SI{0.4}{\degree}$ from the Bragg angle, \emph{i.e.} $\Theta = \Theta_{\mathrm{Bragg}}\pm \Delta \Theta$ and the x-ray beam impinging along the CuMnAs $[110]$ direction for $\Theta = \Theta_{\mathrm{Bragg}}$. For further details see \mbox{reference~\cite{Reimers2022}}.  

\subsection*{Measuring the widths of \SI{180}{\degree} domain walls}
The image processing and data fitting was done using Matlab R2018b. 

To measure the width of the \SI{180}{\degree} domain walls, intensity profiles of linecuts across the walls in the XMLD-PEEM images were extracted and a phenomenological model was fit to the data. For simplicity, the in-plane anisotropy is modelled with a single phenomenological uniaxial term (anisotropy constant $K$). The out-of-plane anisotropy is assumed to be strong enough to keep the N\'eel vector in-plane. In this case orientation of the N\'eel vector is parametrised with a single angular variable $\varphi$ calculated from the easy-axis direction: $\mathbf{n}=(\cos\varphi,\sin \varphi, 0)$. The domain wall profile is then given by a standard expression: 
\begin{equation}\label{eq:180degProfile}
   \varphi - \varphi_\infty = \arccos\left[-\tanh\left(\frac{x-x_0}{d_\mathrm{DW}}\right)\right]\, ,
\end{equation} 
where $\varphi_\infty=0$ or $\pi$ describes the N\'eel vector orientation for $x \longrightarrow - \infty$, $x_0$ is the position of the centre of the domain wall and $d_\mathrm{DW}=\sqrt{A/K}$ is the characteristic lengthscale of the rotation, the ``domain wall widths'', defined by the ratio of the exchange stiffness and anisotropy constants $A$  and $K$. 
Taking into account the functional form of the XMLD-effect one obtains for x-ray polarization along a CuMnAs $[110]$ direction:
\begin{equation}\label{eq:180degXMLDProfile_A}
    MLD_{[110]}  = A_0 \cdot \left[\tanh\left(\frac{x-x_0}{d_\mathrm{DW}}\right)\right]^2 + C \,
\end{equation} 
where $A_0$ and $C$ are constants, which depend on the size of the XMLD signal and on the image normalization. 

To ensure that the profile was obtained in the direction perpendicular to the domain wall, for each point on the domain wall, the angle of the profile was varied, the profiles fitted, with fitting parameters $d_\mathrm{DW}$, $x_0$, $A_0$, $C$,  and the  minimal $d_\mathrm{DW}$ chosen.

A detailed description of the fitting protocols is given in the supporting information.

\subsection*{Simulations}\label{sec:MethodsSim}
For modelling the effect of the elastic incompatibility we use the formalism developed in Refs.~\cite{Wittmann2022,Hendrik2022}. The distribution of the N\'eel vector at the interface with the nonmagnetic substrate is treated as a source of incompatibility charges that create the additional strain $u^\mathrm{elas}$ both in the antiferromagnetic and the nonmagnetic layers.

\subsection*{Domain wall width in CuMnAs stripes}\label{sec:DWModel}
In the presence of additional strain, the domain wall width $d_\mathrm{DW}=\sqrt{A/K_\mathrm{eff}}$ depends on the effective anisotropy $K+H_\mathrm{me}M_{\mathrm{s}}(u^\mathrm{elas}_{xx}-u^\mathrm{elas}_{yy})$, where $H_\mathrm{me}$ is the parameter of magnetoelastic coupling, $M_{\mathrm{s}}/2$ is the sublattice magnetization. 

For a given distribution of the N\'eel vector, the additional strain is calculated as \cite{Wittmann2022} 
\begin{eqnarray}\label{eq_Eshelby_solution}
    u^\mathrm{elas}_{xx}(\mathbf{r})&=&-u^\mathrm{elas}_{yy}(\mathbf{r})=\frac{u_\mathrm{spon}}{4\pi}\int d\mathbf{r}\prime\frac{\cos2\varphi(\mathbf{r}^\prime)}{|\mathbf{r}-\mathbf{r}^\prime|}\delta^\prime(z^\prime),\nonumber\\
\end{eqnarray}
where $u_\mathrm{spon}$ is the value of the spontaneous strain. We used equation~\eqref{eq_Eshelby_solution} to calculate the strain distribution within the CuMnAs stripe assuming a single-domain state $\varphi=0$ within a stripe region. The calculated function $u^\mathrm{elas}_{xx}-u^\mathrm{elas}_{yy}$ is then substituted into the expression for the domain wall width $d_\mathrm{DW}$. We obtain
\begin{equation}\label{eq_DWwidthsCuMnAs}
    \frac{1}{d^2_{DW}} = \frac{1}{D^2} \left( 1-a \mathrm{e}^{ -x / el } \right) \, , 
\end{equation}
where $x$ is the distance from the edge, and $a$, $D$ and $el$ are phenomenological  parameters. $D$ can be interpreted as the domain wall widths at infinity and $el$ as the lengthscale of the problem.
To fit the experimental data with the theoretical curve, we compare the calculated and experimental dependencies $1/d^2_\mathrm{DW}$ as a function of distance from the stripe edge by fitting equation \eqref{eq_DWwidthsCuMnAs} to the experimental data. We obtain $D = \SI{227(8)}{nm}$ and $el = \SI{1.1(2)}{\micro \meter}$.

\subsection*{Domain distribution in Mn$_2$Au stripes}
The equilibrium distribution of the N\'eel vector taking into account incompatibility effects can be calculated by minimization of the destressing energy \cite{Wittmann2022}:
\begin{equation}\label{eq_destressing_energy}
    W_\mathrm{dest}=\frac{1}{2}H_\mathrm{me}M_{\mathrm{s}}u_\mathrm{spon}\int d\mathbf{r}\int d\mathbf{r}^\prime\frac{\cos2\left[\varphi(\mathbf{r}^\prime)-\varphi(\mathbf{r})\right]}{|\mathbf{r}-\mathbf{r}^\prime|}\delta^\prime(z^\prime).
\end{equation}
For the calculation of the domain distribution in Mn$_2$Au we use a multiscale approach (see, \textit{e.g.}\cite{DANIEL20081018,DESIMONE2002283}). We distinguish two principal scales: the large one defined by the stripe (sample) size and an intermediate one, which is much larger than the domain size but much smaller than the sample size. After averaging over the smaller scale $\langle \cos2\varphi(\mathbf{r})\rangle=2\xi(\mathbf{r})-1$, where $\xi(\mathbf{r})$ is the fraction  of the domains with $\varphi=0$, and $\langle\sin2\varphi(\mathbf{r})\rangle=0$, we minimize the destressing energy
\begin{equation}\label{eq_destressing_energy_2}
    W_\mathrm{dest}=\frac{1}{2}H_\mathrm{me}M_{\mathrm{s}}u_\mathrm{spon}\int d\mathbf{r}\int d\mathbf{r}^\prime\frac{(2\xi({x})-1)(2\xi({x}^\prime)-1)}{|\mathbf{r}-\mathbf{r}^\prime|}\delta^\prime(z^\prime),
\end{equation}
assuming that $\xi({x})=1$ at the stripe edges (due to the edge anisotropy).

\medskip

\textbf{Author Contributions}
 SR, MJ, SSD, KWE and PW conceived and led the project.
SSD, KWE and SR devised the XMLD-PEEM imaging with CuMnAs and performed the measurements with FM, OJA, LXB, SFP and PW. 
MJ and SR devised the XMLD-PEEM imaging with \ce{Mn2Au} and performed the measurements with YL, YN and EG.
DC devised the SXDM experiment and performed the measurements with SR, DK, AB, KWE, SSD. DC supervised the data analysis, performed by SR and DK with the help of FK, and coordinated the interpretation of the results.
OG developed the micromagnetic simulations with feedback from MJ, SSD, KWE and SR.
RPC, FK and VN deposited the CuMnAs layers. OJA performed the optical lithography of the samples.
SR, YL and MJ prepared the \ce{Mn2Au} samples, including thin film deposition and lithography.
SR and MJ wrote the manuscript with feedback from all authors. 
% Acknowledgements
\medskip

\textbf{Acknowledgements} \par %delete if not applicable))
The authors thank Diamond Light Source for the allocation of beamtime on beamline I06 under Proposal nos. MM22437-1 and NT27146-1. 
We acknowledge MAX IV Laboratory for beamtime on Beamline NanoMAX under Proposal 20190533 and for time on Beamline MAXPEEM under Proposal 20210863. Research conducted at MAX IV, a Swedish national user facility, is supported by the Swedish Research council under contract 2018-07152, the Swedish Governmental Agency for Innovation Systems under contract 2018-04969, and Formas under contract 2019-02496. 
SR acknowledges support from Diamond Light Source studentship grant STU0201.
SR, MJ and MK acknowledge financial support by the Horizon 2020 Framework, Program of the European Commission under FET-Open Grant No. 863155 (s-Nebula) and funding by the Deutsche Forschungsgemeinschaft (DFG, German Research Foundation) - TRR173/2 - 268565370 Spin-X (Projects A01 and A05). 
OG and  JS acknowledge the Deutsche Forschungsgemeinschaft  via TRR 288 - 422213477 (projects A09) and via TRR 173/2 - 268565370 (projects A03 and B12). 
JS additionally acknowledges funding from Grant Agency of the Czech Republic grant no. 19-28375X. VN and FK acknowledge financial support from the Czech Ministry of Education grants LM2023051 and LNSM-LNSpin, and Czech Science Foundation grant No. 19-28375X.
DK acknowledges the Lumina Quaeruntur fellowship LQ100102201 of the Czech Academy of Sciences and grant number 22-22000M from the Czech Science Foundation. 

\newpage
\begin{suppinfo}

Fitting protocol of the domain wall widths measurement;
Determination of the domain wall width in a non-patterned area of the CuMnAs sample
\end{suppinfo}

\newpage

\bibliography{references}

\end{document}